# Resonant optical spectroscopy and coherent control of $Cr^{4+}$ spin ensembles in SiC and GaN

*William F. Koehl[1,2], Berk Diler[1], Samuel J. Whiteley[3], Alexandre Bourassa[1], N. T. Son[4], Erik Janzén[4], and David D. Awschalom[1,2]*

1. Institute for Molecular Engineering, University of Chicago, Chicago, IL 60637, USA.
2. Materials Science Division, Argonne National Laboratory, Argonne, Illinois 60439, USA.
3. Department of Physics, University of Chicago, Chicago, IL 60637, USA.
4. Department of Physics, Chemistry and Biology, Linköping University, SE-58183 Linköping, Sweden

Spins bound to point defects are increasingly viewed as an important resource for solid-state implementations of quantum information technologies. In particular, there is a growing interest in the identification of new classes of defect spin that can be controlled optically. Here we demonstrate ensemble optical spin polarization and optically detected magnetic resonance (ODMR) of the *S* = 1 electronic ground state of chromium ($Cr^{4+}$) impurities in silicon carbide (SiC) and gallium nitride (GaN). Spin polarization is made possible by the narrow optical linewidths of these ensembles (< 8.5 GHz), which are similar in magnitude to the ground state zero-field spin splitting energies of the ions at liquid helium temperatures. We therefore are able to optically resolve individual spin sublevels within the ensembles at low magnetic fields using resonant excitation from a cavity-stabilized, narrow-linewidth laser. Additionally, these near-infrared emitters possess exceptionally weak phonon sidebands, ensuring that > 73% of the overall optical emission is contained with the defects' zero-phonon lines. These characteristics make this semiconductor-based, transition metal impurity system a promising target for further study in the ongoing effort to integrate optically active quantum states within common optoelectronic materials.

## I. Introduction

Motivated in part by a desire to implement emerging quantum technologies in the solid-state, researchers have become increasingly interested in spins localized at point defects within various electronic and optical materials [1, 2]. Within a subset of these systems, such as rare earth ions [3] and vacancy related complexes in diamond [4 – 6] and silicon carbide [7 – 9], quantum state control and measurement can be accomplished optically. Due to the robust quantum nature of light, the often atom-like optical properties of isolated defects, and the increasing sophistication of modern photonic technologies, such systems have become attractive research testbeds for quantum information science [10 – 16].

Transition metal ion doped solids exhibit a striking range of optical [17], magnetic [18], and electronic [19] behaviors, and are widely used in coherent optics [20] and quantum electronics [21]. This makes them an attractive research target for efforts that seek to develop materials optimized for applications in quantum information. Recent explorations suggest potential roles as single photon emitters [22 – 24], spin-based qubits [25 – 28], and as a storage media for hybrid quantum systems [29]. Chromium ions in the wide-bandgap semiconductors SiC and GaN are reported to exhibit sharp optical transitions in the near-infrared that couple individually to the spin sublevels of the ground state [30, 31]. This suggests a straightforward pathway for optical spin manipulation of these ions, even when many ions are probed simultaneously as an ensemble.

Here we describe resonant optical excitation and microwave coherent control of chromium ions in SiC and GaN. In each ion system, we use a narrow-linewidth laser tuned to the lowest optical transition energy to resonantly excite ions from the ground state into the first excited state. This allows us to directly characterize the optical and spin properties of the ions at liquid helium temperatures and confirm the $S$ = 1 nature of their electronic ground states. Importantly, the ensemble optical linewidths we observe are similar in magnitude to the ions' ground state zero-field spin splitting. Therefore, we are able to spectrally resolve the individual spin sublevels of the ions within these ensembles, even at low magnetic fields (0 – 2500 G). Taking advantage of this feature, we demonstrate optical spin polarization, optically detected magnetic resonance (ODMR), and Rabi driving of spins within these ensembles, establishing that combined optical and microwave methods can be used to initialize, coherently manipulate, and measure spins within this transition metal ion system.

**II. Background**

Previous studies of chromium in SiC and GaN have identified the ion as a substitutional impurity situated on the Si or Ga site of each material, respectively [30 – 35]. Within 4H-SiC, there are two inequivalent Si lattice sites, resulting in two distinct chromium ion species: $Cr_A$ and $Cr_C$. Within GaN there is a single species. At cryogenic temperatures, two sharp optical transitions at 1.1583 eV and 1.1898 eV can be observed in photoluminescence (PL) measurements of chromium-doped 4H-SiC when it is excited non-resonantly with above-bandgap, ultraviolet light (T = 6 K) [30]. These lines correspond to the ZPL optical transitions of $Cr_A$ and $Cr_C$ species respectively when they relax from the first excited state to the ground state. Within GaN, similar measurements reveal that the ZPL energy is located at 1.193 eV (T = 1.8 K) [31]. While the identity of the impurity associated with the line at 1.193 eV is not universally agreed upon [31, 33], the similarity of its optical and spin properties as compared to SiC:Cr and AlN:Cr [34], as well as recent computational results [36, 37], suggest that chromium is the most likely candidate. Very weak phonon sideband emission is observed for all three species, with a small number of discrete phonon modes being clearly visible in the PL signature of each sideband [30, 31].

Under the application of multi-Tesla magnetic fields, these ZPL transitions split according to an $S$ = 1 Zeeman interaction with a $g$-factor ~ 2. Additionally, a doubling of this triplet structure can be observed when the field is aligned parallel to the $c$ axis of the crystal [30]. These and other previous observations were used to conclude that the ground state of chromium ions in these materials is a spin-triplet, orbital singlet ($^3A_2(F)$), while the first excited state is a spin-singlet, orbital doublet ($^1E(D)$). This is the expected electronic structure of a transition metal ion with a $d^2$ electron configuration ($Cr^{4+}$) when it is situated within a strong tetrahedral crystal field environment, as shown in the canonical coordination chemistry energy diagrams developed by Tanabe and Sugano [38]. The ZPL optical transition of each ion species therefore corresponds to an intraconfigurational spin-flip transition, which is consistent with earlier observations of sharp and strain-insensitive optical linewidths, weak phonon sidebands, and relatively long (~ 100 us) optical decay rates [30, 31, 20].

**III. Samples and Setup**

The samples under study include a ~ 5 mm x 8 mm piece of chromium-doped 4H-SiC grown epitaxially on an off-axis, n-type 4H-SiC substrate, as well as a 1.0 cm$^2$ freestanding bulk semi-insulating GaN substrate. The 4H-SiC:Cr$^{4+}$ epilayer was grown via high-temperature chemical vapor deposition (HTCVD) to a thickness of ~ 60 μm with a chromium density of $10^{15} – 10^{16}$ cm$^{-3}$. The GaN sample is 468 μm thick and was grown via hydride vapor phase epitaxy (HVPE) as a commercial substrate by Kyma, Inc. (part

No. GB.SE.010.DSP.D). The GaN sample is compensation-doped with $Fe^{3+}$ to pin the Fermi level near mid-gap, and the chromium ions observed in our studies are unintentional dopants introduced during the growth process. Other GaN samples purchased from Kyma with nominally identical specifications exhibit varying amounts of $Cr^{4+}$ luminescence, depending on the growth batch.

The majority of our data are derived from photoluminescence excitation (PLE) measurements. In these measurements, we collect only the phonon sideband emission of the $Cr^{4+}$ ions as they relax from the first excited state. Excitation into the first excited state is accomplished using a narrow-linewidth, tunable laser tuned to resonance with the ZPL optical transition. In all PLE measurements, the laser spot diameter at the sample surface is ~ 30 μm with excitation powers of 5 – 10 mW. Two-color excitation experiments are performed by generating optical sidebands on the laser with a phase-modulating electro-optic modulator (EOM). Samples are mounted in a liquid helium flow cryostat with microwave and optical access, and a motorized permanent magnet mount located behind the cryostat is used to generate ≤ 2500 G magnetic fields along the *c* axis of the sample. A short-terminated microwave coplanar waveguide situated beneath the sample enables microwave driving of the $Cr^{4+}$ ion spins. Detection of phonon sideband emission is accomplished using either an InGaAs spectrometer or an InGaAs femtowatt photoreceiver. For time-resolved measurements of spin and optical dynamics, a digital delay generator is used to handle pulse timing of optical and microwave pulses. Amplitude modulation of the laser is achieved with an acousto-optic modulator (AOM), while a microwave switch with a switching time of ~5 ns is used to modulate the microwave driving field.

In Figs. 1a,c we also report photoluminescence (PL) measurements in which the entire emission spectrum (both the ZPL and phonon sideband) is recorded as ions relax from the first excited state. In these measurements, optical excitation is accomplished using a Ti:sapphire laser tuned to an energy far above the first excited state (710 nm/1.74 eV). Data are normalized using spectra taken with a calibrated white light source installed in place of the sample and no laser excitation, in order to compensate for wavelength-dependent absorption in the collection optics. Detailed schematics and an extended description of the experimental setup for both the PL and PLE configurations can be found in the Supplementary Information.

**IV. Resonant Optical Excitation**

In Fig. 1 we compare the results of PL and PLE measurements on both the 4H-SiC and GaN samples at liquid helium temperatures and B = 0 G. For each ion species, a single ZPL and associated sideband can be observed in the PL data (Figs. 1a,c), consistent with results from previous literature [30, 31]. As discussed in the Supplementary Information, we find that the fraction of luminescence emitted within the ZPL features is 75% for 4H-SiC:Cr ions and 73% for GaN:Cr.

Sharp increases in PLE signal at 1.1587 eV and 1.1898 eV (4H-SiC, Fig. 1b) and 1.193 eV (GaN, Fig. 1d) match the ZPL energies observed in PL measurements of the same samples, demonstrating that we are able to resonantly excite all three forms of chromium ion with the narrow-linewidth, tunable laser. Fine frequency scans centered at these energies are shown in Figs. 1e-g. While only a single maximum is resolved in the PLE signal for 4H-SiC:$Cr_A$ ions, two maxima are clearly observed for both the 4H-SiC:$Cr_C$ and GaN:Cr impurities. In these two latter cases, we fit the data to the sum of two Lorentzians:

$$PLE(f) = \frac{A}{\pi}\left(\frac{\left(\frac{\Gamma}{2}\right)}{(f-f_0)^2 + \left(\frac{\Gamma}{2}\right)^2}\right) + \frac{B}{\pi}\left(\frac{\left(\frac{\Gamma}{2}\right)}{(f-f_1)^2 + \left(\frac{\Gamma}{2}\right)^2}\right) + C$$

Where $f_0$ and $f_1$ (A and B) are the central frequencies (amplitudes) of the two Lorentzians, Γ is the full-width half maximum (FWHM) linewidth of both Lorentzians, and C is a constant to account for non-zero background offset in the signal. For the 4H-SiC:$Cr_C$ defect, we find that the linewidth of these features is Γ = 7.42 ± 0.07 GHz at 30 K and 0 G, with an energy splitting between the two maxima of $(f_0 - f_1)$ = 6.46 ± 0.05 GHz. For the GaN:$Cr^{4+}$ ions, these values are Γ = 8.28 ± 0.14 GHz and $(f_0 - f_1)$ = 6.91 ± 0.05 GHz. We note that in the case of 4H-SiC:$Cr_C$, the energy splitting between the two peaks is roughly similar to the previously reported values of 6.70 GHz [32] and 6.0 GHz [30] given for the ground state zero-field spin splitting. This suggests that these two peaks correspond to the $m_S$ = 0 and $m_S$ = ± 1 spin sublevels of the ion's purported S = 1 ground state, and we tentatively label them as such for ease of description in the measurements that confirm this identification detailed below. For the remainder of this paper, we will focus primarily on data collected from the 4H-SiC:$Cr_C$ impurity, but similar results have been observed for the other two ion species and are given in the Supplementary Information.

### V. Optical Spin Polarization and Magnetic Field Dependence

To confirm that the two peaks observed within the $Cr_C$ ZPL correspond to spin sublevels, we first study the effect of magnetic field when it is applied along the *c* axis of the crystal at 30 K (Fig. 2a). We see that the PLE lineshape indeed changes as the magnetic field is increased from 12 – 2500 G, although the exact nature of this evolution is somewhat obscured by the inhomogeneous broadening of the optical transitions. By converting the data into a differential measurement in which the data at B = 13 G is subtracted from each PLE scan (Fig. 2b), we see more clearly what occurs. As the magnetic field is applied, a dip flanked symmetrically on either side by two small peaks forms. This dip is centered at the same frequency as the feature labeled $m_S$ = ±1 in Fig. 1f, and grows in magnitude as the field is increased. Similarly, the two peaks on either side grow and appear to move slowly outward away from the dip. This is the expected behavior for an optical transition between an excited state spin singlet and a ground state spin triplet; under the application of a magnetic field, the degenerate optical transitions connecting the $m_S$ = ±1 sublevels of the ground state to the singlet excited state will begin to split apart in energy according to the Zeeman effect. Note that no signal related to the $m_S$ = 0 sublevel is observed in the differential data since its energy remains unchanged by the magnetic field.

We can characterize this magnetic field dependent behavior more precisely at lower temperatures using optical spin polarization. According to the level structure of Ref. [30] and Fig. 3b, selective optical excitation of one ground state spin sublevel with a narrow-linewidth laser will pump the system into another sublevel via resonant excitation followed by spontaneous emission. A polarized ion will then remain dark and inaccessible to the laser until a spin-flip occurs. At temperatures below ~ 20 K, the spin-lattice relaxation time $T_1$ of the $Cr_C$ ions becomes much longer than the optical relaxation time $T_{opt}$ (Figs. 2d-e and Supplementary Information). This results in a substantially reduced PLE signal at these lower temperatures (Fig. 2c), as well as long-lived optical spin polarization within the sub-ensemble of ions excited by the laser. A recovery of luminescence, however, should be observed if both spin sublevels are excited simultaneously.

To test this assumption, we perform two-color experiments on the defect ensemble at T = 15 K and B = 0 G. We set the laser frequency $f_0$ to the frequency of the $m_S$ = 0 peak in Fig. 1f. We then add optical sidebands to the laser emission at $f_S = f_0 \pm f_{EOM}$ by modulating the EOM with a microwave signal in the range of 0 – 10 GHz. If we are indeed polarizing the impurity spins through resonant optical excitation at $f_0$, then an increase in PLE should be observed when $f_{EOM} = D$, the zero-field spin splitting of the ground state. As shown in Fig. 3a, this is in fact what we observe. Fits to the data reveal that $D$ = 6.711 ± 0.001 GHz, which is consistent with the splitting measured between the two PLE maxima of Fig. 1f, as well as with the reported values given in Refs. [30] and [32]. Additional features observed at $f_{EOM}$ = 3.37 and 2.24 GHz are measurement artifacts resulting from the second and third order sideband harmonics generated by the EOM. We note that, because the spin sublevels of the ground state can be separately resolved in the optical spectra, we are generating a net spin polarization within the ion ensemble, rather than simply three equal populations of ions polarized into each of the sublevels (see Supplementary Information).

Importantly, the linewidth of the feature at 6.711 GHz is much narrower (489 ± 5 MHz) than the PLE linewidths observed in single-color experiments at 30 K. This is because we probe only a subpopulation of the impurity ensemble in these two-color measurements, so that the resulting linewidth is set by factors such as the laser linewidth, laser stability, and ion spectral diffusion [39]. This narrow linewidth allows us to very clearly observe the 6.711 GHz feature splitting in two as a magnetic field is once again applied along the crystal c axis (Fig. 3c). Using fits to this data, we plot the evolution of this splitting as a function of magnetic field strength (Fig. 3d). A Zeeman relationship consistent with an $S$ =1 system is clearly apparent, with a linear fit revealing a $g$-factor of 2.01 ± 0.05. Identical studies of 4H-SiC:$Cr_A$ and GaN:Cr support the $S$ = 1 nature of these ions as well, with respective $D$ values of 1.057 ± 0.001 GHz and 7.335 ± 0.002 GHz (See Supplementary Information).

**VI. Optically Detected Magnetic Resonance & Microwave Coherent Control**

An increase in the PLE signal at low temperatures will also occur if we apply microwave radiation resonant with the ground state spin splitting energy. This enables ODMR of the optically polarized spin ensemble, which can then be exploited to probe the coherent time-dynamics of the spin ensemble. In Fig. 4a we excite the sample at T = 15 K with a single optical frequency tuned to the center of the $m_S$ = 0 optical transition of Fig. 1f. We then apply continuous microwave excitation to the sample while scanning the microwave frequency between 0 – 10 GHz. At zero magnetic field, a single resonance is observed at $f$ = 6.707 GHz. In the plot, PLE contrast is defined as ΔPLE = ($I_{ON} - I_{OFF}$)/$I_{OFF}$, where $I_{ON}$ ($I_{OFF}$) is the integrated intensity of the phonon sideband emission when the microwave driving field is on (off). Therefore, the intensity of the phonon sideband roughly doubles at B = 0 G under these driving conditions.

As a magnetic field is applied, the resonance splits into two roughly equal peaks as expected for an $S$ = 1 Zeeman system, with a fit yielding a $g$-factor of 2.018 ± 0.004 (Fig. 4b). These observations are identical to those measured using two-color optical excitation, and further confirm that we have been probing the ground state spin of this system. A fit to data taken at lower microwave power to reduce power broadening reveals a linewidth of 8.6 ± 0.5 MHz (Fig. 4c). This corresponds to an inhomogeneous spin coherence time of $T_2^* = 1/(\pi \cdot \Gamma)$ = 37 ± 2 ns. We note, however, that the sample under study is heavily doped with chromium ions. Assuming a homogeneous distribution of ions, the average interspin distance is roughly 50 – 100 nm. This may introduce broadening due to dipolar spin-spin interactions

[40] or local strain inhomogeneity within the sample [41, 42], and suggests that improvements in materials quality and a more dilute spin system may lead to increases in observed coherence. ODMR studies of 4H-SiC:$Cr_A$ and GaN:Cr yield similar results, although a ~ 27x larger linewidth is observed for GaN:Cr as expected due to interactions with the surrounding nuclear spin bath (see Supplementary Information).

The inhomogeneous coherence time inferred from our frequency domain measurements is much shorter than the maximum rate at which our current experimental setup can drive a π rotation. This limitation precludes us from measuring $T_2^*$ or $T_2$ in the time domain using Ramsey and spin echo techniques. However, as seen in Fig. 4d, we are still able to generate and detect Rabi oscillations in the spin ensemble by detuning the microwave driving field from resonance. In this regime, the Rabi frequency $\omega_\rho$ is determined largely by the driving field detuning $\Delta\omega$, rather than the driving power [43]. With a microwave detuning of $\Delta\omega$ = 40 MHz, we observe coherent driving at a Rabi frequency of $\omega_\rho$ = 41.7 ± 0.5 MHz. The dependence of $\omega_\rho$ on detuning, as well as Rabi data for $Cr_A$, are given in the Supplementary Information. While a large microwave detuning prevents full population inversion by restricting the spins to a small polar region of the Bloch sphere, these measurements still demonstrate as a proof of principle the coherent control and time-resolved optical spectroscopy of chromium ions in SiC. With extensions of $T_2^*$ and fabrication of on-chip microwave sources for stronger driving, continued explorations of dynamical properties should be possible.

**VII. Summary and Outlook**

In this study we have shown that $Cr^{4+}$ ion spins in SiC and GaN can be directly manipulated with a narrow-linewidth laser tuned to resonance with the first excited state ZPL optical transition. The ensemble optical linewidths we observe are similar in magnitude to the zero-field spin splitting of the electronic ground state, which enables ensemble optical spin polarization and measurement at liquid helium temperatures and low magnetic fields. This capability was used to precisely determine the *g*-factor and zero-field spin splitting *D* of all three spin species, as well as demonstrate coherent Rabi driving and time-resolved optical spectroscopy of the two $Cr^{4+}$ species in 4H-SiC.

Importantly, most of the luminescence from the first excited state is contained within the narrow ZPL optical transitions, which form a very simple lambda structure with no competing intermediate transitions other than those of the weakly coupled phonon sideband. The ions therefore exhibit quite atom-like optical properties, while also being embedded within semiconductor hosts amenable to advanced optoelectronic device design. This suggests their use as quantum emitters that couple efficiently to chip-scale, integrated photonic control structures – an effort that depends critically on limiting both intrinsic and extrinsic sources of optical and non-radiative loss [44].

It is intriguing to note that while the observed optical transitions are protected from the influence of strain and phonons, the ground state spin still couples readily to mechanical degrees of freedom. For instance, local trigonal distortions in the crystal lattice surrounding an ion are capable of generating sizeable increases in the zero-field spin splitting energy [45]. This effect underlies the large difference in *D* observed for $Cr_C$ and $Cr_A$ ions in 4H-SiC [30]. Furthermore, acoustically driven magnetic resonance studies of comparable transition metal ion systems have demonstrated large spin-strain couplings that exceed those of popular vacancy-related defects [46 – 48, 15]. These properties, in conjunction with the piezoelectric nature of SiC and GaN, imply that $Cr^{4+}$ ions may be a promising platform for mechanical manipulation of quantum information.

Future improvements in materials quality should improve key optical and spin properties explored in this study. Improvements in strain homogeneity should result in narrower optical linewidths, as observed for similar spin systems in ruby and MgO [49, 50]. This would enable an even higher degree of optical spin polarization than observed here, especially if spin sublevels become fully separated within the optical spectra (see Supplementary Information). Ensemble ODMR linewidths may also narrow due to reductions in strain-related inhomogeneous broadening of the spin resonance. Additionally, a lower spin density could extend coherence times by improving magnetic isolation of dopant ions. For example, spin coherence times ranging from microseconds up to ~ 1 ms have been observed in other, more well-isolated transition metal ion systems [26, 28, 51, 52].

Lastly, we note that the knowledge gained in our studies of $Cr^{4+}$ ions in SiC and GaN is transferrable to other materials systems exhibiting comparable optical and spin characteristics. Similar magnetic ions are found in other common semiconductors and remain largely unexplored [53 – 55]. Small molecules that reproduce the basic structural characteristics of the ions studied here can be synthesized [56 – 57] and would enable fine tuning of the local quantum environment via ligand design. Our results are therefore applicable to a range of transition metal ion systems, and present exciting opportunities in the ongoing effort to exploit defect-localized spins for explorations in quantum science and engineering.

### Acknowledgements

We thank B. B. Zhou, D. J. Christle, C. G. Yale, F. J. Heremans, P. J. Mintun, G. Wolfowicz, E. C. Vincent, and C. P. Anderson for advice on the manuscript, and A. L. Falk for arranging sample transfer. W.F.K. and D.D.A. were supported by the U.S. Department of Energy, Office of Science, Office of Basic Energy Sciences, Materials Sciences and Engineering Division. B.D., S.J.W., and A.B. were supported by the Air Force Office of Scientific Research, the Army Research Office, and the National Science Foundation DMR-1306300. E.J. and N.T.S. acknowledge support from the Linköping Linnaeus Initiative for Novel Functional Materials (LiLi-NFM) and the Knut and Alice Wallenberg Foundation.

### References

1) D. D. Awschalom, L. C. Bassett, A. S. Dzurak, E. L. Hu, and J. R. Petta, Quantum Spintronics: Engineering and Manipulating Atom-Like Spins in Semiconductors, Science **339**, 1174 (2013).
2) J. R. Weber, W. F. Koehl, J. B. Varley, A. Janotti, B. B. Buckley, C. G. Van de Walle, and D. D. Awschalom, *Quantum computing with defects*, Proc. Natl Acad. Sci. USA **107**, 8513 (2010).
3) J. H. Wesenberg, K. Mølmer, L. Rippe, and S. Kröll, Scalable designs for quantum computing with rare-earth-ion-doped crystals, Phys. Rev. A **75**, 012304 (2007).
4) F. Jelezko, T. Gaebel, I. Popa, A. Gruber, and J. Wrachtrup, *Observation of coherent oscillations in a single electron spin*, Phys. Rev. Lett. **92**, 076401 (2004).
5) B. Pingault, J. N. Becker, C. H. H. Schulte, C. Arend, C. Hepp, T. Godde, A. I. Tartakovskii, M. Markham, C. Becher, and M. Atatüre, *All-Optical Formation of Coherent Dark States of Silicon-Vacancy Spins in Diamond*, Phys. Rev. Lett. **113**, 263601 (2014).
6) L. J. Rogers, K. D. Jahnke, M. H. Metsch, A. Sipahigil, J. M. Binder, T. Teraji, H. Sumiya, J. Isoya, M. D. Lukin, P. Hemmer, and F. Jelezko, *All-Optical Initialization, Readout, and Coherent Preparation of Single Silicon-Vacancy Spins in Diamond*, Phys. Rev. Lett. **113**, 263602 (2014).
7) W. F. Koehl, B. B. Buckley, F. J. Heremans, G. Calusine, and D. D. Awschalom, *Room temperature coherent control of defect spin qubits in silicon carbide*, Nature **479**, 84 (2011).

8) D. J. Christle, A. L. Falk, P. Andrich, P. V. Klimov, J. Ul Hassan, N. T. Son, E. Janzén, T. Ohshima, and D. D. Awschalom, *Isolated electron spins in silicon carbide with millisecond coherence times*, Nature Mater. **14**, 160 (2015).
9) M. Widmann, S.-Y. Lee, T. Rendler, N. T. Son, H. Fedder, S. Paik, L.-P. Yang, N. Zhao, S. Yang, I. Booker, A. Denisenko, M. Jamali, S. A. Momenzadeh, I. Gerhardt, T. Ohshima, A. Gali, E. Janzén and J. Wrachtrup, *Coherent control of single spins in silicon carbide at room temperature*, Nature Mater. **14**, 164 (2015).
10) M. Zhong, M. P. Hedges, R. L. Ahlefeldt, J. G. Bartholomew, S. E. Beavan, S. M. Wittig, J. J. Longdell, and M. J. Sellars, *Optically addressable nuclear spins in a solid with a six-hour coherence time*, Nature **517**, 177 (2015).
11) Andreas Walther. 'Coherent processes in rare-earth-ion-doped solids'. PhD thesis, Division of Atomic Physics, LTH (2009).
12) B. Hensen, H. Bernien, A. E. Dréau, A. Reiserer, N. Kalb, M. S. Blok, J. Ruitenberg, R. F. L. Vermeulen, R. N. Schouten, C. Abellán, W. Amaya, V. Pruneri, M. W. Mitchell, M. Markham, D. J. Twitchen, D. Elkouss, S. Wehner, T. H. Taminiau, and R. Hanson, *Loophole-free Bell inequality violation using electron spins separated by 1.3 kilometres*, Nature **526**, 682 (2015).
13) C. G. Yale, F. J. Heremans, B. B. Zhou, A. Auer, G. Burkard, and D. D. Awschalom, *Optical manipulation of the Berry phase in a solid-state spin qubit*, Nature Phot. **10**, 184 (2016).
14) A. Sipahigil, K. D. Jahnke, L. J. Rogers, T. Teraji, J. Isoya, A. S. Zibrov, F. Jelezko, and M. D. Lukin, *Indistinguishable Photons from Separated Silicon-Vacancy Centers in Diamond*, Phys. Rev. Lett. **113**, 113602 (2014).
15) A. L. Falk, P. V. Klimov, B. B. Buckley, V. Ivády, I. A. Abrikosov, G. Calusine, W. F. Koehl, Á. Gali, and D. D. Awschalom, *Electrically and Mechanically Tunable Electron Spins in Silicon Carbide Color Centers*, Phys. Rev. Lett. **112**, 187601 (2014).
16) P. V. Klimov, A. L. Falk, D. J. Christle, V. V. Dobrovitski, and D. D. Awschalom, *Quantum entanglement at ambient conditions in a macroscopic solid-state spin ensemble*, Sci. Adv. **1**, e1501015 (2015).
17) B. Henderson, G. F. Imbusch, *Optical Spectroscopy of Inorganic Solids* (Clarendon Press, Oxford, 1989) pp. 408 – 444.
18) A. Abragam and B. Bleaney, *Electron Paramagnetic Resonance of Transition Ions* (Clarendon Press, Oxford, 1970).
19) K. A. Kikoin and V. N. Fleurov, *Transition Metal Impurities in Semiconductors: Electronic Structure and Physical Properties* (World Scientific, Singapore, 1984).
20) C. E. Webb and J. D. C. Jones (eds.) *Handbook of Laser Technology and Applications Volume II: Laser Design and Laser Systems, Volume 2* (Institute of Physics Publishing, Bristol, 2004) pp. 307 – 352.
21) T. Dietl and H. Ohno, *Dilute ferromagnetic semiconductors: Physics and spintronic structures*, Rev. Mod. Phys. **86**, 187 (2014).
22) T. Gaebel, I. Popa, A. Gruber, M. Domhan, F. Jelezko, and J. Wrachtrup, *Stable single-photon source in the near infrared*, New J. Phys. **6**, 98 (2004).
23) G. D. Marshall, T. Gaebel, J. C. F. Matthews, J. Enderlein, J. L. O'Brien, and J. R. Rabeau, *Coherence properties of a single dipole emitter in diamond*, New J. Phys. **13**, 055016 (2011).
24) I. Aharonovich, S. Castelletto, B. C. Johnson, J. C. McCallum, D. A. Simpson, A. D. Greentree, and S. Prawer, Chromium single-photon emitters in diamond fabricated by ion implantation, Phys. Rev. B **81**, 121201(R) (2010).

25) K. A. Baryshnikov, L. Langer, I. A. Akimov, V. L. Korenev, Yu. G. Kusrayev, N. S. Averkiev, D. R. Yakovlev, and M. Bayer, *Resonant optical alignment and orientation of $Mn^{2+}$ spins in CdMnTe crystals*, Phys. Rev. B **92**, 205202 (2015).
26) J. Tribollet, J. Behrends, and K. Lips, *Ultra long spin coherence time for $Fe^{3+}$ in ZnO: A new spin qubit*, Eur. Phys. Lett. **84**, 20009 (2008).
27) R. E. George, J. P. Edwards, and A. Ardavan, *Coherent Spin Control by Electrical Manipulation of the Magnetic Anisotropy*, Phys. Rev. Lett. **110**, 027601 (2013).
28) J. M. Zadrozny, J. Niklas, O. G. Poluektov, and D. E. Freedman, *Millisecond Coherence Time in a Tunable Molecular Electronic Spin Qubit*, ACS Cent. Sci. **1**, 488 (2015).
29) D. I. Schuster, A. P. Sears, E. Ginossar, L. DiCarlo, L. Frunzio, J. J. L. Morton, H. Wu, G. A. D. Briggs, B. B. Buckley, D. D. Awschalom, and R. J. Schoelkopf, *High-Cooperativity Coupling of Electron-Spin Ensembles to Superconducting Cavities*, Phys. Rev. Lett. **105**, 140501 (2010).
30) N. T. Son, A. Ellison, B. Magnusson, M. F. MacMillan, W. M. Chen, B. Monemar, and E. Janzén, *Photoluminescence and Zeeman effect in chromium-doped 4H and 6H SiC,* J. Appl. Phys. **86**, 4348 (1999).
31) R. Heitz, P. Thurian, I. Loa, L. Eckey, A. Hoffmann, I. Broser, K. Pressel, B. K. Meyer, and E. N. Mokhov, *Identification of the 1.19-eV luminescence in hexagonal GaN*, Phys. Rev. B **52**, 16508 (1995).
32) P. G. Baranov, V. A. Khramtsov, and E. N. Mokhov, *Chromium in silicon carbide: electron paramagnetic resonance studies*, Semicond. Sci. Technol. **9**, 1340 (1994).
33) J. Baur, K. Maier, M. Kunzer, U. Kaufmann, J. Schneider, H. Amano and I. Akasaki, T. Detchprohm, and K. Hiramatsu, *Infrared luminescence of residual iron deep level acceptors in gallium nitride (GaN) epitaxial layers*, Appl. Phys. Lett. **64**, 857 (1994).
34) J. Baur, U. Kaufmann, M. Kunzer, J. Schneider, H. Amano, I. Akasaki, T. Detchprohm, and K. Hiramatsu, *Photoluminescence of residual transition metal impurities in GaN*, Appl. Phys. Lett. **67**, 1140 (1995).
35) K. Pressel, R. Heitz, L. Eckey, I. Loa, P. Thurian, A. Hoffmann, B. K. Meyer, S. Fischer, C. Wetzel, and E. E. Haller, *Identification of Transition Metals in GaN*, Mat. Res. Soc. Symp. Proc. **395**, 491 (1996).
36) U. Gerstmann, M. Amkreutz, and H. Overhof, *Paramagnetic Defects*, Phys. Stat. Sol. (b) **217**, 665 (2000).
37) W.-C. Zheng, S.-Y. Wu, and J. Zi, *Theoretical studies of the g-shift for $Cr^{4+}$ ions in GaN crystal from crystal-field and charge-transfer mechanisms*, J. Phys.: Condens. Matter **13**, 7459 (2001).
38) Y. Tanabe and S. Sugano, *On the Absorption Spectra of Complex Ions II*, J. Phys. Soc. Jpn. **9**, 766 (1954).
39) W. E. Moerner (ed.) *Persistent Spectral Hole-Burning: Science and Applications* (Springer-Verlag, 1988)
40) A. L. Falk, B. B. Buckley, G. Calusine, W. F. Koehl, V. V. Dobrovitski, A. Politi, C. A. Zorman, P. X.-L. Feng, D. D. Awschalom, *Polytype control of spin qubits in silicon carbide*, Nature Comm. **4**, 1819 (2013).
41) A. M. Stoneham, *Shapes of Inhomogeneously Broadened Resonance Lines in Solids*, Rev. Mod. Phys. **41**, 82 (1969)
42) R. Calvo, M. C. G. Passeggi, and C. Fainstein, Phys. Lett. **37A**, 201 (1971).
43) C. P. Slichter, Principles of Magnetic Resonance (Springer-Verlag, Berlin, 1996).
44) W. B. Gao, A. Imamoglu, H. Bernien, and R. Hanson, *Coherent manipulation, measurement and entanglement of individual solid-state spins using optical fields*, Nature Phot. **9**, 363 (2015).

45) K. Žďánský, *Strain-Induced Zero-Field Splitting of d3 and d8 Ions in Cubic Crystals*, Phys. Rev. **159**, 201 (1967).
46) Z. Wen-Chen, *Theoretical calculations of the spin-lattice coupling coefficients $G_{11}$ and $G_{44}$ for MgO:$Ni^{2+}$ crystals*, Phys. Rev. B **40**, 7292 (1989).
47) J. F. Clare and S. D. Devine, *Measurement of the spin-strain coupling tensor in ruby by ultrasonically-modulated EPR*, J. Phys. C: Solid State Phys. **13**, 865 (1980).
48) J. Teissier, A. Barfuss, P. Appel, E. Neu, and P. Maletinsky, *Strain Coupling of a Nitrogen-Vacancy Center Spin to a Diamond Mechanical Oscillator*, Phys. Rev. Lett. **113**, 020503 (2014).
49) P. E. Jessop and A. Szabo, *High Resolution Measurements of the Ruby $R_1$ Line at Low Temperatures*, Optics Communications **33**, 301 (1980).
50) G. F. Imbusch, W. M. Yen, A. L. Schawlow, G. E. Devlin, and J. P. Remeika, *Isotope Shift in the R Lines of Chromium in Ruby and MgO*, Phys. Rev. **136**, A481 (1964).
51) M. Warner, S. Din, I. S. Tupitsyn, G. W. Morley, A. M. Stoneham, J. A. Gardener, Z. Wu, A. J. Fisher, S. Heutz, C. W. M. Kay, G. Aeppli, *Potential for spin-based information processing in a thin-film molecular semiconductor*, Nature **503**, 504 (2013).
52) K. Bader, M. Winklera, and J. van Slageren, *Tuning of molecular qubits: very long coherence and spin–lattice relaxation times*, Chem. Commun. **52**, 3623 (2016).
53) S. Gabilliet, V. Thomas, J. P. Peyrade, and J. Barru, *The Luminescence at 0.795 eV from GaAs:Nb: A Zeeman Spectroscopy*, Phys. Lett. A **119**, 197 (1986).
54) D. Ammerlahn, B. Clerjaud, D. Côte, L. Köhne, M. Krause, and D. Bimberg, *Spectroscopic Investigation of Neutral Niobium in GaAs*, Mater. Sci. Forum **258-263**, 911 (1997).
55) P. Thurian, I. Loa, P. Maxim, A. Hoffmann, C. Thomsen, and K. Pressel, *The $V^{3+}$ center in AlN*, Appl. Phys. Lett. **71**, 2993 (1997).
56) B. K. Bower and H. G. Tennent, *Transition Metal Bicyclo[2.2.1]hept-1-yls*, J. Amer. Chem. Soc. **94**, 2512 (1972).
57) W. Mowat, A. J. Shortland, N. J. Hill, and G. Wilkinson, *Elimination Stabilized Alkyls. Part II. Neopentyl and Related Alkyls of Chromium(IV)*, J. Chem. Soc., Dalton Trans. 770 (1973).

**Figure Captions**

**Figure 1:** (a) Photoluminescence (PL) spectrum of Cr doped 4H-SiC at T = 30 K. Zero phonon line (ZPL) optical emission from two distinct ensembles of $Cr^{4+}$ ions are visible at 1.1587 and 1.1898 eV. These two ion species are labeled $Cr_A$ and $Cr_C$, respectively, and lie at structurally inequivalent silicon sites in the crystal lattice. Phonon sideband emission at lower energy is magnified 25x for clarity. (b) Photoluminescence excitation (PLE) spectrum of Cr doped 4H-SiC at T = 30 K. Data taken by collecting phonon sideband emission while scanning a narrow line laser across the first excited state ZPL energies of $Cr_A$ and $Cr_C$. (c) PL spectrum of Cr doped GaN at T = 30 K. A sharp ZPL at 1.193 eV is visible with a small phonon sideband at lower energy. Sideband is magnified 25x. (d) PLE spectrum of Cr doped GaN at T = 30 K. Data taken in same manner as panel b. Fine PLE scans demonstrating direct excitation of $Cr^{4+}$ ions at T = 30 K are shown in (e) 4H-SiC:$Cr_A$ (f) 4H-SiC:$Cr_C$ and (g) GaN:Cr. Fits to the data are shown in the latter two panels, where spin sublevels of the electronic ground state can be resolved optically at zero magnetic field.

**Figure 2:** (a) PLE spectrum of 4H-SiC:$Cr_C$ as a function of magnetic field applied along the crystal *c* axis (T = 30 K). (b) The effect of the magnetic field on the PLE spectrum is clarified by subtracting data at B = 13 G from spectra taken at higher fields. With increasing magnetic field, a single dip forms at the energy

corresponding to the $m_S = \pm 1$ feature of Fig. 1f. Simultaneously, two peaks emerge on either side while moving symmetrically outward. This is the behavior expected from the proposed spin singlet to spin triplet optical transition. (c) PLE spectra of 4H-SiC:$Cr_C$ as a function of temperature. Spectra taken at T = 20 K and below multiplied 5x for clarity. (d) Optical decay of 4H-SiC:$Cr_C$ from the first excited state at T = 20 K. Ions are excited non-resonantly with 710 nm light and a monochromator in the collection path is used to collect only the $Cr_C$ ZPL emission. A fit to the data reveals an optical decay time $T_{opt}$ = 145.6 ± 6.1 μs. (e) $T_1$ time of 4H-SiC:$Cr_C$ spins as a function of temperature. Note that $T_1$ is ~ 45x longer than $T_{opt}$ near 20 K. Error bars are 95% confidence intervals.

**Figure 3:** (a) Two-color excitation of 4H-SiC:$Cr_C$ ions at T = 15 K and B = 0 G. $Cr_C$ spins are polarized into the $m_S = \pm 1$ sublevels via resonant optical excitation of the $m_S = 0$ sublevel. A recovery of the PLE signal is observed when a second laser line at $f_{EOM}$ = 6.711 ± 0.001 GHz is used to simultaneously excite the $m_S = \pm 1$ optical transition. (b) Schematic of the $Cr^{4+}$ level structure and two-color resonant optical excitation. (c) Two-color excitation experiment as a function of magnetic field applied along the crystal *c* axis (T = 15 K). The feature at 6.711 GHz splits in two as expected for a spin triplet. (d) Fits to the peaks in panel c reveal a clear Zeeman relationship with *g* = 2.01 ± 0.05. Error bars are 95% confidence intervals. Features centered at $f_{EOM}$ = 3.37 and 2.24 GHz in panels a and c are artifacts due to higher order optical sidebands generated by the EOM.

**Figure 4:** (a) Optically detected magnetic resonance (ODMR) of 4H-SiC:$Cr_C$ spins as a function of magnetic field applied along the crystal *c* axis at T = 15 K. Spins are polarized into the $m_S = \pm 1$ sublevels via resonant excitation of $m_S = 0$ optical transition seen in Fig. 1f. When $\omega_{RF}$ is resonant with the spin splitting energy, the PLE signal increases. Small oscillations seen at the base of the signal are artifacts due to imperfect transmission in the microwave driving lines. (b) Fits to ODMR peaks as a function of magnetic field reveal a clear Zeeman relationship consistent with an *S* = 1 spin system with *g* = 2.018 ± 0.004. Error bars are 95% confidence intervals. (c) Low power ODMR measurement at T = 15 K and B = 289 G with limited power broadening yields a linewidth of 8.6 ± 0.5 MHz. This corresponds to an inhomogeneous spin coherence time $T_2^*$ = 37 ± 2 ns. (d) Rabi driving of $Cr_C$ spins at T = 15 K and B = 289 G with a driving field detuning Δω = 40 MHz. A fit to the data reveals a Rabi decay time $T_\rho$ = 24.2 ± 1.9 ns.

# Figures

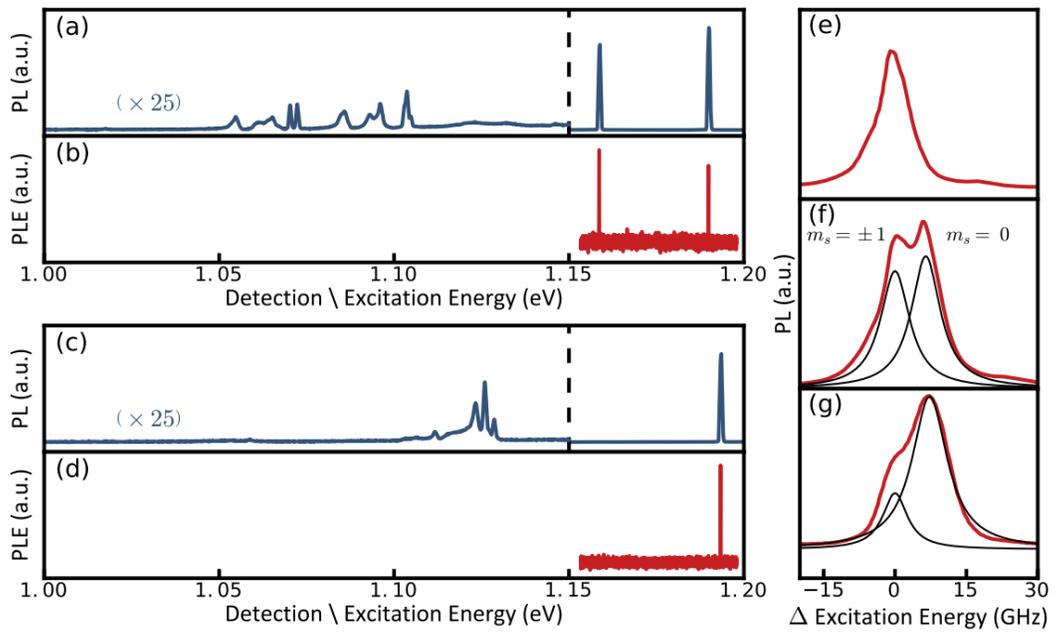

**Figure 1**

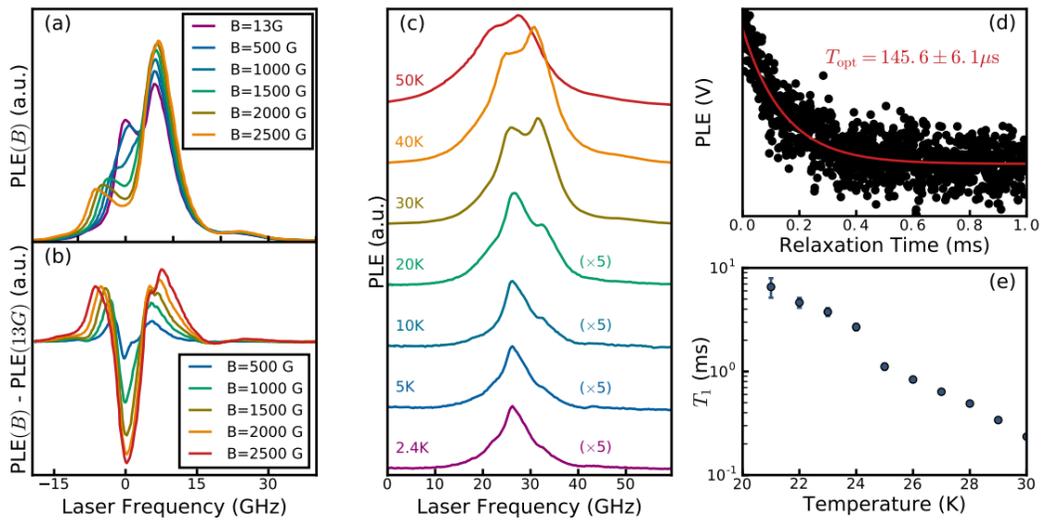

**Figure 2**

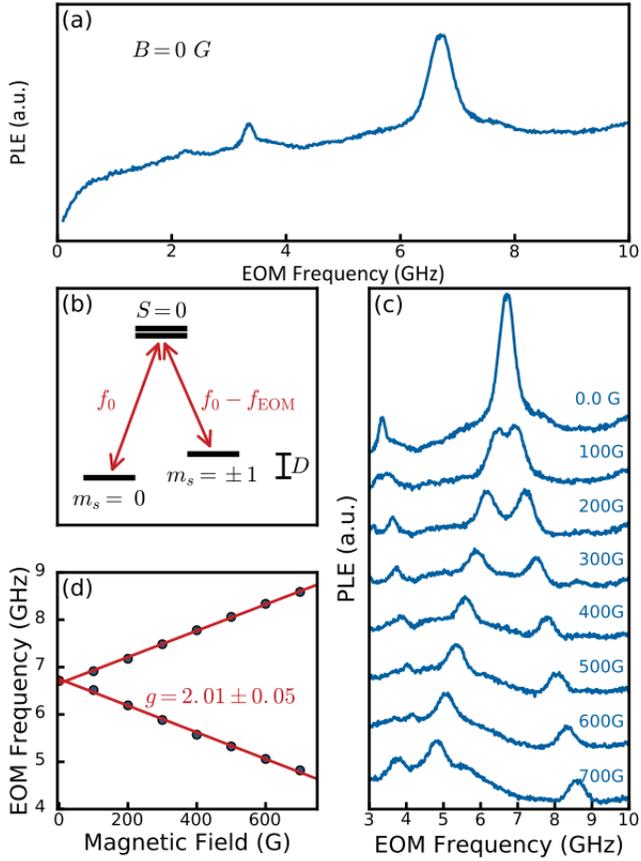

**Figure 3**

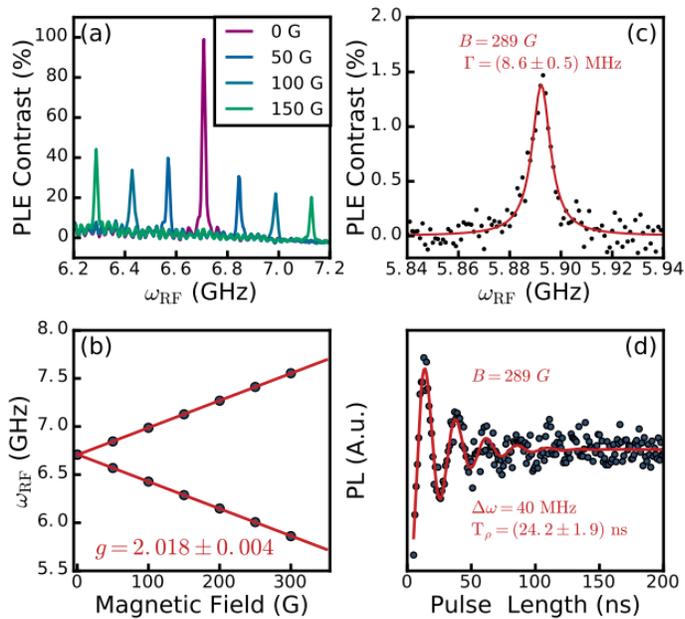

**Figure 4**